\documentclass[aip,jap,preprint]{revtex4-1}
 
\usepackage{bm}
\usepackage{graphicx}   
\usepackage[hang]{subfigure}

\begin{document}

\title {Current Amplification with Vertical Josephson Interferometers}
\thanks{Submitted to Journal of Applied Physics}

\author{R. Monaco}
\affiliation{Istituto di Cibernetica 'E. Caianiello' del CNR, Comprensorio Olivetti, 80078  Pozzuoli, Italy}
\email{roberto@sa.infn.it}

\date{\today}
\begin{abstract}
It has long been recognized that a control current $I_a$ injected into the section of a two-junction superconducting quantum interference device (SQUID) is able to produce a change of its critical current $I_c$, so that a current gain $g=\left|dI_c/dI_a\right|$ can be identified. We investigate the circumstances under which large gains can be achieved by using vertical Josephson interferometers which are characterized by small loop inductances. We discuss the theory of operation of such a novel device, its performances and its advantages with respect to planar interferometers used in the previous works. Two potential applications are addressed.
\end{abstract}

\maketitle
\tableofcontents
%

\section{INTRODUCTION}

Josephson interferometers lie at the core of the most sensitive detectors of magnetic flux currently available\cite{handbook}. They are amazingly versatile, being able to measure any physical quantity that can be converted to a flux. As demonstrated by Clarke and Paterson\cite{clarke71a} three decades ago, one can use a dc SQUID as a current amplifier by injecting a signal current $I_a$ into part of the superconducting loop and detecting the resultant change in the critical current $I_c$. The current gain $\left| dI_c /dI_a \right|$ can be increased by making the inductance in the two branches of the interferometer asymmetric, thereby skewing the transfer function $I_c(I_a)$\cite{tesche}. At liquid helium temperature current gains as large as $20$ were demonstrated with a $Pb-–Cu/Al-–Pb$ SQUID although in a small current range\cite{clarke71a}; however, due to the Nyquist noise, at liquid nitrogen temperature, the largest achievable gains are of the order of few units\cite{koelle,kleiner}. 


\noindent In this work we propose to replace the planar interferometer used in all the previous investigations\cite{clarke71a,koelle,kleiner} with a Vertical Josephson Interferometer (VJI) consisting of two superconducting strips of width $w$ separated by an insulator layer with thickness $t_{ox}$ shunted by two Josephson Tunnel Junctions (JTJs) a distance $l$ apart. This configuration is strongly reminiscent of that used by Jaklevic\cite{jaklevic} {\it et al.} in their pioneering work on quantum interference effects in Josephson tunneling in which the pickup loop was formed by two $Sn$ films separated by a Formvar resin spacer approximately $1 \mu m$ thick. Recently the same geometrical configuration has been revived by Granata\cite{granata} {\it et al.} to realize a controllable flux transformer in quantum computation applications. However, in none of the cited works was the VJI employed as a current sensor or amplifier.

\begin{figure}[tb]
\subfigure[ ]{\includegraphics[width=14cm]{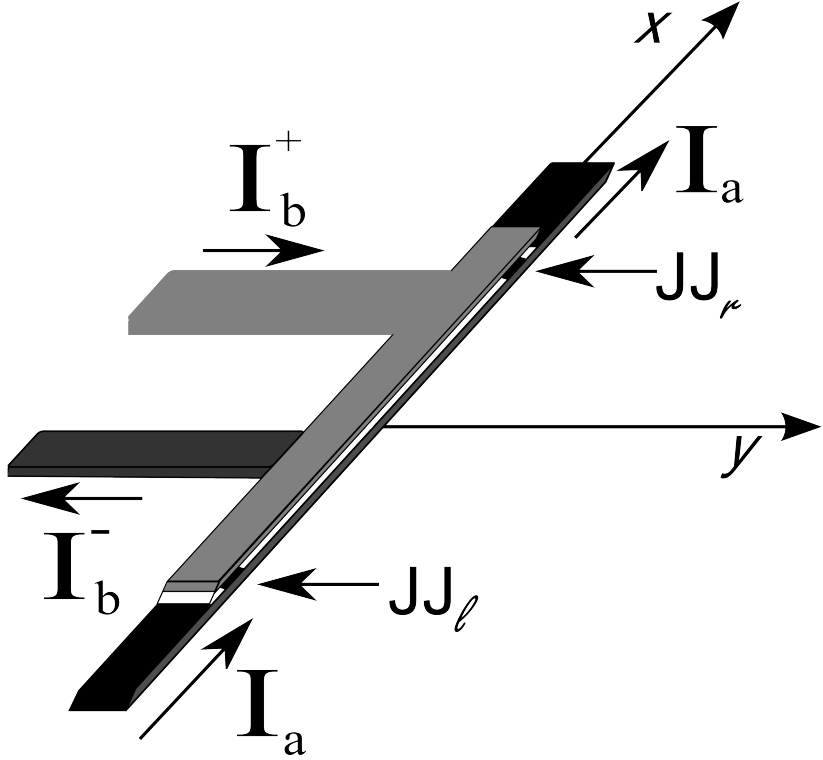}}
\caption{Three-dimensional sketch of a superconducting amplifier: the current $I_a$ flowing in the bottom thin film (in black) modulates the critical current of the built-on vertical Josephson interferometer.}
\label{amplifier}
\end{figure}

\noindent The three-dimensional sketch of the proposed device is shown in Fig.\ref{amplifier}, together with the coordinate system used in this work. Here the current $I_a$ flows along the base electrode of a parallel plate superconducting-insulating-superconducting strip-line terminated at each extremity by a properly sized hysteretic JTJ; the insulating layer can be made of a anodic native oxide or some silicon oxides or a combination of both (it should not be confused with the much thinner Josephson tunneling barrier). The JTJs can be resistively shunted, to transduce the current modulation into voltage changes. If the thicknesses $t_s$ of the two superconducting strips forming the VJI are much greater than their superconducting penetration depth $\lambda$, then the magnetic thickness is $d=t_{ox}+2\lambda$, and the magnetic flux $\Phi_a$ associated with the current $I_a$ flowing in any strip is efficiently coupled to the VJI pickup loop area $A_{loop}=ld$ of the order of $100\mu m^2$, i.e., much smaller than the typical area of planar loops. Small pickup loop areas correspond to small loop inductances which make the interferometer relatively insensitive to external magnetic disturbance\cite{tesche}. In addition, the device full compatibility with any all-refractory Niobium technology developed for the fabrication of window-type JTJs permits to  continuously adjust its electrical and geometrical parameters over a wide range. Furthermore, it is easy to design since it does not require input coils, feedback loops, flux transformers and so on. For the same reasons a VJI also requires a very limited space and it is ideal for highly integrated environments. It is worth to anticipate that the presented findings do not constitute an improvement in the state-of-the-art of current sensors, but are intended for specific applications where the device simplicity matters more than its ultimate performances.

\noindent The paper is organized in the following way. The basic equations describing a Josephson interferometer are listed in Sec.II together with a short review of the principle of current amplification with dc SQUIDS. Then our VJI-based current amplifier is described (Sec.III): we discuss the working principles, estimate the main electrical  characteristics (such as current gain, sensitivity, dynamic range, etc.), discuss the impact of the device geometry on its current gain and provide hints for the design of real devices. This is followed by a discussion of two possible applications in Sec.IV. Finally, the conclusion are drawn Sec.V.

\section{THE DOUBLE-JUNCTION INTERFEROMETER}

\begin{figure}[tb]
\includegraphics[width=14cm]{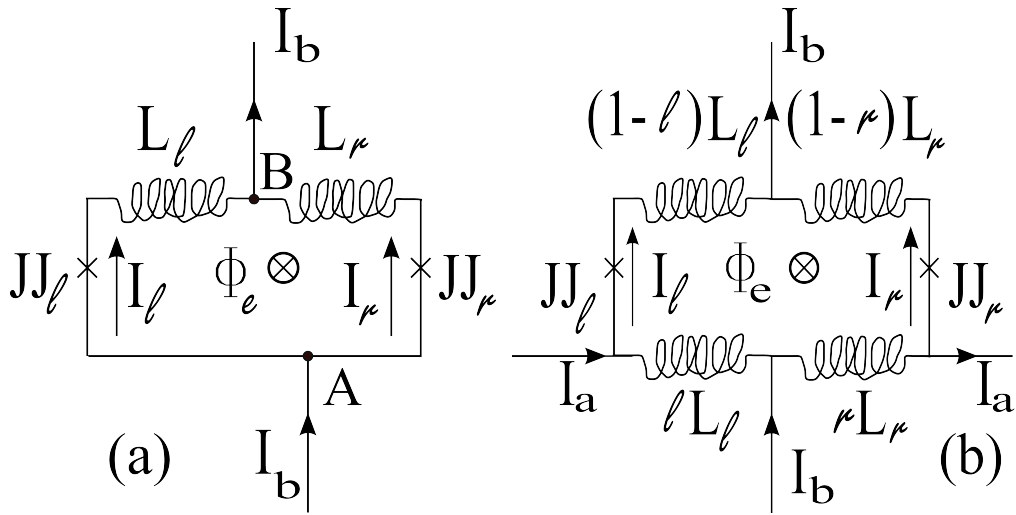}
\caption{(a) Circuit diagrams for: (a) a two-junction interferometer and (b) a  four-terminal device consisting of a double junction interferometer having a current $I_a$ flowing in its lover arm, which necessitate partitioning the inductances. The junctions are maked by an $\times$.}
\label{schemes}
\end{figure}

Let us consider the double junction interferometer, shown in Fig.\ref{schemes}(a), consisting of a superconducting loop linked to an external magnetic flux $\Phi_e$ and interrupted by two Josephson links marked by an $\times$. The direct bias current $I_b$ flowing through the junctions can be fed into the loop at an arbitrary point $A$ and extracted out at an arbitrary point $B$ which makes, in general, the interferometer lacking of any feed symmetry. Throughout this paper it is assumed that the system thermal energy $k_B T$ is much lower than the Josephson energies $\Phi_0 I_{c\ell,r}/2\pi$, so that thermal fluctuations can be neglected and a deterministic analysis is possible\cite{tesche}; in practice, at liquid $He$ temperature, the low temperature limit requires that $I_{c\ell,r}>1 \mu A$. The bias current $I_b$ can be expressed in terms of the gauge invariant phase differences $\phi_\ell$ and $\phi_r$ across the left and right junctions, respectively, as\cite{barone}:

\begin{equation}
I_b(\phi_\ell, \phi_r)= I_\ell(\phi_\ell)+I_r(\phi_r)=I_{c\ell} \sin \phi_\ell +I_{cr} \sin \phi_r.
\label{Ib}
\end{equation}

\noindent In this work we will neglect the junction capacities since we limit our interest to stationary effects, although high frequency applications can be thought of up to the $GHz$ range\cite{muck}. The single-valuedness condition for the phase of the superconducting wave function around the loop (fluxoid quantization) yields\cite{barone}:

\begin{equation}
n \Phi_0=\frac{\phi_\ell- \phi_r}{2\pi}\Phi_0 +\Phi_e + L_\ell I_\ell - L_r I_r ,
\label{quantize0}
\end{equation}

\noindent in which $n$ is an integer number called {\it winding} number and $\Phi_0=h/2e \simeq 2.07 \times 10^{-15}Wb$ is the magnetic flux quantum. In the above equation $L_\ell$ and $L_r$ are positive coefficients having units of inductance and $L=L_\ell + L_r$ is the loop self-inductance. When $I_{c\ell}=I_{cr}$ and  $L_\ell = L_r$, one talks of a symmetric interferometer configuration. It is useful to rewrite Eq.(\ref{quantize0}) in terms of normalized quantities and with the explicit dependence on the Josephson phases $\phi_{\ell,r}$:

\begin{equation}
2\pi n=\phi_\ell- \phi_r +2\pi \phi_e + \pi \beta_\ell \sin \phi_\ell - \pi \beta_r \sin \phi_r,
\label{quantize1}
\end{equation}

\noindent with $\phi_e=\Phi_e/\Phi_0$ and $\beta_{l,r}=2 L_{\ell,r} I_{c\ell,r} /\Phi_0$. The $\beta$ parameters measure in each interferometer's arm the ratio between the magnetic energy in the self-inductance $L_{\ell,r}I_{c\ell,r}^2/2$ and the Josephson coupling energy $\Phi_0 I_{c\ell,r}/2\pi$ ($\beta_{\ell,r}=\pi/2$ when the ratio is one). Postulating that the applied magnetic flux $\Phi_e$ does not affect in any way the junction supercurrents, then the largest zero-voltage $I_b$ value versus the external flux $\phi_e$ can be numerically and graphically computed for different values of the parameters $I_{c\ell}$, $I_{cr}$, $\beta_\ell$, $\beta_r$ and of the winding number $n$\cite{tsang,peterson}; normalizing currents to $I_0=(I_{c\ell}+I_{cr})/2$, the interferometer positive and negative critical currents, respectively, $i_c^+(\phi_e)=\max_{\phi_\ell,\phi_r} I_b(\phi_\ell,\phi_r)/I_0$ and $i_c^-(\phi_e)=\min_{\phi_\ell,\phi_r} I_b(\phi_\ell,\phi_r)/I_0$, are oscillatory functions with the unitary period.

\begin{figure}[tb]
\includegraphics[width=14cm]{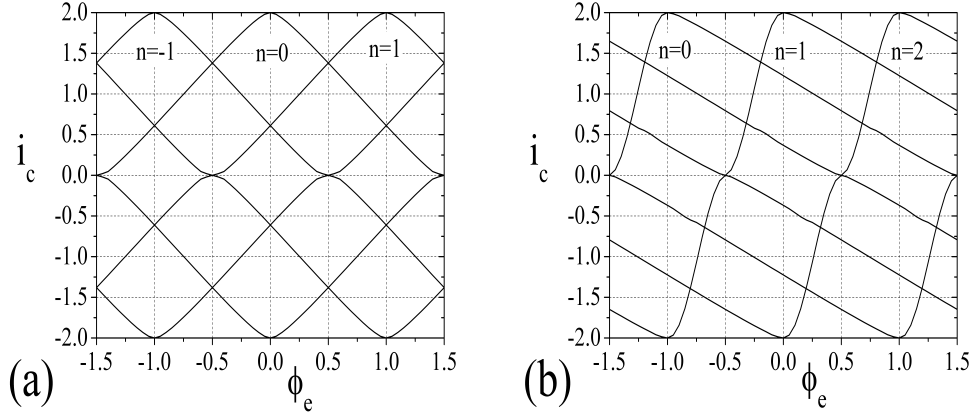}
\caption{Numerically computed positive and negative interferometer threshold curves $i_c(\phi_e)$ with $I_{c\ell}=I_{cr}$ and $\beta_\ell+\beta_r=2$: (a) symmetric case, $\beta_\ell=\beta_r=1$, and (b) (fully) asymmetric case $\beta_\ell=0$ and $\beta_r=2$. In the experiments only the stable solutions are measured, which correspond to the upper and lower envelopes of the threshold curves.}
\label{plots}
\end{figure}

\noindent Figs.\ref{plots} shows the interferometer magnetic diffraction patterns (also called threshold curves) $i_c(\phi_e)$ in two cases with equal junction critical currents ($I_{c\ell}=I_{cr}$). In Fig.\ref{plots}(a) we consider the symmetric case with $\beta_\ell=\beta_r=\beta_0=1$, characterized by even and symmetric threshold curves: $i_c^-(\phi_e)=-i_c^+(\phi_e)$. For $\beta_0>1$ the threshold curves are piecewise linear with the absolute slope decreasing with $\beta_0$ as $2\beta_0^{-1}-\beta_0^{-2}$. This behavior is at variance with the asymmetric case ($\beta_\ell \neq \beta_r$) in which the threshold curve tends to be tilted retaining the symmetry with respect to the simultaneous inversion of the current and of the magnetic flux:  $i_c^-(\phi_e)=-i_c^+(-\phi_e)$; this effect becomes stronger the more $\beta_\ell$ differs from $\beta_r$. Tesche and Clarke\cite{tesche} showed that not only inductance asymmetries, but also asymmetries in critical current of the two Josephson junctions lead to a skewing of $i_c$ vs $\phi_e$ and to a shift from the origin $\phi_{o}$ of the maxima positions. As a result of the skewness, the slope $\left| di_c/d\phi_e \right|$  can be large over a certain $\phi_e$ range which provides a better sensitivity to flux changes. Particularly interesting is the case of ultimate asymmetry, i.e., when one $\beta$ parameter is vanishingly small; the case with $\beta_\ell=0$ and $\beta_r=2$ is shown in Fig.\ref{plots}(b) (in this paper we set $\beta_\ell \leq \beta_r$). Now the threshold curve envelope is made by linearly decreasing segments with slope $-2(\beta_r^{-1}-\beta_r^{-2})$ alternated by steeply increasing S-shaped branches whose appearance is practically independent on the $\beta_r$, except that for an offset term. The analysis of the numerical solutions led to the empirical expression:

\begin{equation}
i_c(\phi_e) \simeq i_{cr}+ i_{c\ell} \sin 2\pi \left( \frac{\phi_e}{1+2\beta_\ell}+\phi_{o}\right),
\label{ic}
\end{equation}


\noindent with $\phi_{o}=\beta_r/2-n +1/4$. 

\noindent It is well known that the double junction modulation depth $i_{c,min}^{+}/i_{c,max}^{+}$ is independent on the interferometer feed asymmetry\cite{fulton,tsang}. In fact, it only  depends on $\beta_L= (2L/\Phi_0) \min \left\{ I_{c\ell},I_{cr} \right\}$: when $\beta_L < 1$, the SQUID has a pronounced modulation depth, i.e., $i_{c}^{min} << i_{c}^{max}=2$, while, $i_{c}^{min}/i_{c}^{max} = 1 - \beta_L^{-1} + (2/\pi) \beta_L^{-3/2}+ O(\beta_L^{-2})$, in the opposite case\cite{kleiner}. From Eq.(\ref{ic}) we can calculate the slope at $i_c=i_{c}^{min}$; to the first order of approximation in $(1- i_{c}^{min}/i_{c}^{max})$ we have:

$$\left|\frac{di_c}{d\phi}\right|_{i_{c}^{min}}  \simeq \frac{4 \pi}{(1+2\beta_\ell)} \sqrt{\frac{2 \alpha}{1+\alpha} \left( 1 - \frac{i_{c}^{min}}{i_{c}^{max}} \right) } \simeq  $$

\begin{equation}
\simeq \frac{4 \pi}{(1+2\beta_\ell)} \sqrt{\frac{2 \alpha}{1+\alpha}\left( \frac{1}{\beta_L} - \frac{2}{\pi \beta_L^{3/2}}\right)} \simeq  \frac{4 \pi}{(1+2\beta_\ell)} \sqrt{\frac{2 \alpha}{1+\alpha}\frac{1}{\beta_L}},
\label{slope}
\end{equation} 

\noindent where $\alpha=I_{c\ell}/I_{cr}$. We will make use of this equation in next section to estimate the current gain of a asymmetric dc-SQUID. 


\indent Let us now consider the case when a direct current $I_a$ is fed to the bottom interferometer arm, as illustrated by the  electrical scheme in Fig.\ref{schemes}(b): here, one must know what fraction of the total inductance is coursed by the control current which, in the most general fashion, lead to partitioned inductances with $\ell ,r \in [0,1]$\cite{peterson}. To analyze this circuit, we begin by observing that the additional magnetic flux $\Phi_a$ associated with $I_a$ is $\Phi_a = L_a I_a$, where $L_a$ is given by the sum of the bottom inductances: $L_a=\ell L_{\ell}+r L_{r}$. Next, while Eq.(\ref{Ib}) still holds true, the fluxoid quantization around the loop Eq.(\ref{quantize0}) now requires that:

$$ n\Phi_0=\frac{\phi_\ell- \phi_r}{2\pi}\Phi_0 +\Phi_e +$$ 
$$(1-\ell)L_{\ell}I_\ell+ \ell L_{\ell}(I_\ell-I_a)-
(1-r)L_{r}I_r - rL_{r}(I_r+I_a), $$

\noindent or, rearranging, 

\begin{eqnarray}
2\pi n= \phi_\ell- \phi_r +2\pi (\phi_e - \phi_a) +  \nonumber \\
\pi \beta_\ell \sin \phi_\ell - \pi \beta_r \sin \phi_r.
\label{quantize2}
\end{eqnarray}

\noindent We note that Eq.(\ref{quantize2}) equals Eq.(\ref{quantize1}), if the effective normalized flux through the loop is $\phi = \phi_e - \phi_a$\cite{peterson}. In other words, the signal current $I_a$ can be detected and measured through the change of the magnetic flux threading the loop area; in the absence of an externally applied magnetic field ($\phi_e=0$), the linked flux is proportional to $I_a$ with the proportionality constant depending on the geometrical details of the device through the inductance $L_a$.

\section{THE VERTICAL JOSEPHSON INTERFEROMETER}

\noindent Without any practical loss of generality, we will only consider those VJIs having $\ell=r=1/2$ meaning that the bias current $I_b$ enters and leaves the interferometer at the same abscissa leaving full arbitrarity on $L_{\ell,r}$.

\noindent To determine the device's performances we first have to compute the transfer function $\Phi_a(I_a)$; being $\Phi_a=L_a I_a$, the inductance $L_a$ has to be found. Insofar as the strip-line magnetic thickness $d$ is much smaller than the width $w$ of the current-carrying strip, the magnetic field $H_a$ perpendicular to the loop area generated by a current $I_a$ flowing along the $x$-direction is approximately given by $I_a/w$\cite{swihart}; in fact, in the wide strip approximation, most of the magnetic energy is confined in the region between the plates and the fringing field can be ignored. As the strip width becomes narrower, the fringe field effects become more important and may dominate if $w$ and $d$ are comparable\cite{chang}. Then, to a first approximation, the flux linked to the loop is $\Phi_a=\mu_0 A_{loop} H_a =\mu_0 l d  H_a ={\mathcal L}_0l I_a$, where we have introduced the strip-line inductance per unit length ${\mathcal L}_0=\mu_0 d/w$\cite{orlando,faris,granata} which also takes into account the kinetic inductance, due to the motion of superelectrons. Indeed, the inductance of a superconducting strip transmission line was analytically derived by Chang\cite{chang} as far as the strip linewidth $w$ exceeds about the insulation thickness $t_{ox}$; in the thick film approximation ($t_s>>\lambda$), its formula for the inductance per unit length reduces to:

\begin{equation}
{\mathcal L}={\mathcal L}_0/K(w/d,t_s/d), 
\label{L}
\end{equation}

\noindent where the fringing-field factor $K$ lowers the inductance value being always larger than unity. Ultimately we have that $L_a={\mathcal L}l$. With a similar reasoning it is found that, for a VJI, the total loop inductance is $L=L_\ell+L_r=2L_a$, as intuitively expected. Inserting experimentally reasonable values $w=10d=20t_s=5\mu m$ (corresponding to $K\simeq1.4$), we obtain ${\mathcal L}\simeq 10^{-7} H/m=0.1 pH/\mu m$. Choosing, for instance $l=200\mu m$, it is $L_a \simeq 20pH$ and the $I_c(I_a)$ periodicity $\Phi_0/L_a$ is about $100 \mu A$. Such periodicity corresponds, in the symmetric configuration, to a current responsivity of the SQUID (defined as the variation of the critical current as function of the external magnetic flux variation) of $50 \mu A/\Phi_0$. 

\noindent With $I_c= i_c I_0$ and $I_a= \phi_a \Phi_0/L_a$, the current gain $g=\left|dI_c/dI_a\right|$ is given by $\beta_a \left|di_c/d\phi\right| /2$ where $\beta_a = 2L_a I_0/\Phi_0$. Both $\beta_a$ and $di_c/d\phi$ depend on the specific device geometry; Figs.\ref{linear}(a) and (b) show the top views of two such geometries. In the first case the configuration is fully symmetric, so that $L_{l} = L_{r} = {\mathcal L} l=L_a$, i.e., $\beta_\ell = \beta_r= \beta_0= \beta_a$. Since, as previously seen, in this case the threshold curve has a triangular shape with absolute slope - to the lowest order - equal to $2/\beta_0=2/\beta_a$, then, independently on electrical and geometrical parameters of the interferometer, $g \simeq 1$ indicating a one-to-one correspondence between the applied signal and the critical current change. This is at variance with the values predicted in Ref.\cite{clarke71a} ($g \simeq \pi$) and in Ref.\cite{kleiner} ($g \simeq 2$) for a symmetric interferometer. Such small gains although not attractive for the realization of amplifiers can be nevertheless sufficient to sense currents otherwise difficult to detect.

\noindent In the second case the bias current $I_b$ is injected asymmetrically in such a way that $L_{\ell}=0$ and $L_{r} = 2{\mathcal L} l = 2L_a=L$ (i.e., $\beta_\ell=0$ and $\beta_r = 2\beta_a$), so that the response $I_c(I_a)$ will be skewed as that in Fig.\ref{plots}(b) and tends towards a saw-tooth waveform. On the (less steep) linear branches, with $\left|di_c/d\phi\right| \simeq 2/\beta_r= 1/\beta_a$, the current gain is even lower: $g \simeq 0.5$; fortunately, large current gains can be achieved flux biasing the interferometers on the S-shaped regions where the threshold curve is steepest. To maximize the gain, a proper static flux $\phi_e$ has to be applied so as to bias the SQUID at the steepest point of the $I_c(I_a)$ curve; with a VJI this can be easily achieved by a controlled magnetic field applied normal to loop area, i.e., in the $y$-direction of Fig.\ref{amplifier}. Considering that, if $\alpha= I_{c\ell} / I_{cr} \leq 1$, $\beta_L = 4 \alpha \beta_a / (1+\alpha)$, then Eq.(\ref{slope}) allows us to derive the following general expression for the current gain:

\begin{figure}[tb]
\includegraphics[width=14cm]{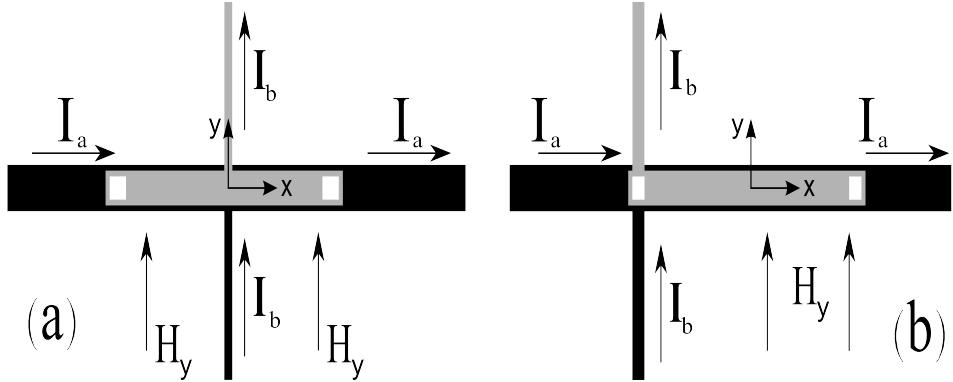}
\caption{Sketch (not in scale) of two linear current amplifier based on vertical Josephson interferometers. The base electrode is in black, the top electrode is in gray and the junction area is white: (a) symmetric bias and (b) asymmetric bias.}
\label{linear}
\end{figure}

\begin{equation}
g(\beta_\ell,\beta_r) = \frac{ \pi \sqrt{\beta_r}}{(1+2\beta_\ell)}.
\label{guadagno}
\end{equation}

\noindent In other words, for practical implementations as a current amplifier, an interferometer asymmetry as large as possible is  required\cite{clarke71a}. However what matters are the absolute $\beta$ values and not their ratio. In the experiments it is sometime difficult to realize devices having a very high degree of inductance asymmetry; when this is the case,  $\beta_\ell$ can be kept small by properly reducing $I_{c\ell}$. As it is, Eq.(\ref{guadagno}) does not contains any dependence on the critical current asymmetry and it also applies when $I_{c\ell} \neq I_{cr}$; a small correction to the gain can be derived from Eq.(\ref{slope}) when $\alpha \neq 1$. Considering the numbers already used above and with $2I_0=(I_{c\ell}+I_{cr})\approx 500 \mu A$, then $\beta_r \simeq 10$, i.e., current gain as large as $10$ can be readily achieved.





\noindent The efficient operation of the amplifier requires that the applied current $I_a$ essentially flows in the bottom arm of the interferometer, so that the fraction $\rho$ of the current $I_a$ inductively diverted in the top arm through the Josephson links can be neglected. According to the notations of figure Fig.\ref{schemes}(b), $\rho$ is given by: 

$$ \rho=\frac{\ell L_{\ell}+ r L_{r}}{L_{\ell}+L_{r}+ L_{jl}+L_{jr}}=\frac{L_a}{2L_a+ L_{jl}+L_{jr}},$$

\noindent with the last equality inferred from considering that, for a VJI, $\ell L_{\ell}+ r L_{r}= (1-\ell)L_{\ell}+(1+r)L_{r}=L_a$. The condition $\rho<<1$ is satisfied when the sum of the non-linear inductances of the Josephson elements $L_{j\ell,r}= \Phi_0/2\pi I_{c\ell,r} \cos \phi_{\ell,r}$ is much larger than $L_a$ which is certainly true whenever the interferometer is operated close to its critical currents where, as Eq.(\ref{slope}) indicates, $\cos \phi_{\ell,r} \approx 0$ for at least one JTJ. It can be shown that the magnetic flux $\Phi_a$ associated to the control current $I_a$ is reduced by a factor $1-2\rho$, for $\rho<<1$.

\section{PRACTICAL APPLICATIONS}

It has been recognized that SQUID-based devices theoretically can operate at very high frequencies up to several tens of gigahertz making them attractive for several applications, such as high-frequency amplifiers, oscillators, or phase shifters\cite{muck}. More specifically vertical interferometers are also currently used as double junction SIS mixers for wide-band millimeter and sub-millimeter-waves receivers\cite{noguchi}. Hitherto we analyzed the d.c. properties of the VJIs and in this section we will consider a couple of its possible static applications.

\subsection{Supercurrent detection}

Nowadays, the superconducting electronics is invariably based on thin films. Because of its purely inductive internal resistance, the most natural application of the device under study is to sense the zero-voltage current flowing through a superconducting film in a sort of {\it clamp meters}. This task is accomplished by simply building a properly designed VJI on top of the current carrying strip. The optimal sensor design stems from a compromise between sensitivity and dynamic range. The current sensitivity of the detector, set by the smallest change in $I_c$ that we can appreciate, among other things, is proportional to the amplifier gain which, as seen in the previous section, for an asymmetric interferometer increases with $\beta_a$. Since $\beta_a$, in turn, is proportional to the VJI length $l$, any large gain can be achieved in principle; however, the largest $\beta_a$ value is set by the current range $\Delta I_a$ over which current amplification is required. Indeed, reminding that in the large inductance limit, to the lowest order, $\Delta I_c=I_{c,max} - I_{c,min}\simeq (\Phi_0/2L)(I_{c\ell}+I_{cr})/\min \left\{ I_{c\ell},I_{cr} \right\}= (1+\alpha) \Phi_0/2L$, then $\Delta I_a= I_0 \Delta i_a =I_0 \Delta i_c /g \simeq I_0 (2/\beta_T)/(2\pi \sqrt{\beta_a})= I_0 /\pi \beta_L \sqrt{\beta_a}$. Being $\beta_L=4 \alpha \beta_a/(1+\alpha)$, after some algebra, we end up with $ \Delta I_a= \sqrt{(1+\alpha)/(8 \pi \alpha I_{cr})}(\Phi_0/{\mathcal L l})^{3/2}$ from which the interferometer length $l$ can be inferred, once the critical currents $I_{cr}$ and $I_{c\ell}=\alpha I_{cr}$ have been established from independent criteria.  

\noindent A right critical current $I_{cr}=500\mu A$ can be obtained from a $5\times 5 \mu m^2$ window-type planar JTJs with low-temperature Josephson current density $J_c \simeq 20 \mu A/\mu m^2= 2kA/cm^2$. If needed, it is possible to trim $I_{cr}$ to a somehow smaller value by rising the setup temperature or by applying a local magnetic field. $Nb$-based junctions fabricated with such electrical and geometrical parameters would have a Josephson penetration depth larger than their linear dimension and a critical magnetic field $H_c=\Phi_0/(2 \mu_0 \lambda_{Nb}l_J )\simeq 2 kA/m$, with the $Nb$ London depth $\lambda_{Nb}\simeq 100nm$ and $l_j= 5 \mu m$, while the magnetic field values involved in the device operation is of the order of $50\mu A/5 \mu m = 10 A/m $, i.e., way too small to affect the junctions critical currents.

\noindent We conclude this section with the observation that, if a higher current sensitivity is required, then the distance $l$ between the JTJs can be increased, but correspondingly the junction critical current $I_{cr}$ has to be consistently decreased in order to keep the product $l^{3} I_{cr}$ constant. We like to stress that, because of its limited sensitivity, our device cannot compete with any SQUID based galvanometer\cite{grana} or current comparators\cite{ccc} used nowadays. To give a practical example, considering, as in the previous section, $I_c \approx 500\mu A$ and $g \approx 10$ and assuming that the measurement resolution of the critical current is about $1$ part in $10^3$, we have a signal resolution of about $50 nA$. It is worth to mention that Russo {\it et al.}\cite{russo} have recently demonstrated that at liquid $He$ temperatures the interferometer critical current can be determined with at least one order of magnitude better accuracy by measuring the switching current distributions.



\subsection{Nondestructive fluxoids readout}

\noindent As one more possible application of the proposed device, let us consider the layout made by two superconducting loops biased as depicted in Fig.\ref{fluxmeter}(a), whose corresponding electrical diagram is shown in Fig.\ref{fluxmeter}(b). We will see that it is possible to use the upper loop consisting of a vertical double junction interferometer to detect the persistent current associated to the eventual flux quanta trapped in the lower loop whose geometry is uninfluential. Several factors conspire to render detection of the persistent currents extremely difficult; the current flows only around a closed path, so the effect is lost if a device like an ammeter is put into the circuit to measure it directly. The vortex imaging is nowadays accomplished by means of high-sensitivity SQUID microscopy techniques in which a small area pick up coil scans the ring neighborhood\cite{minami,kirtley}.

\noindent The system analysis now requires two fluxoid quantization conditions; for the VJI loop threaded by an external flux $\Phi_e$ it is:

\begin{figure}[tb]
\includegraphics[width=14cm]{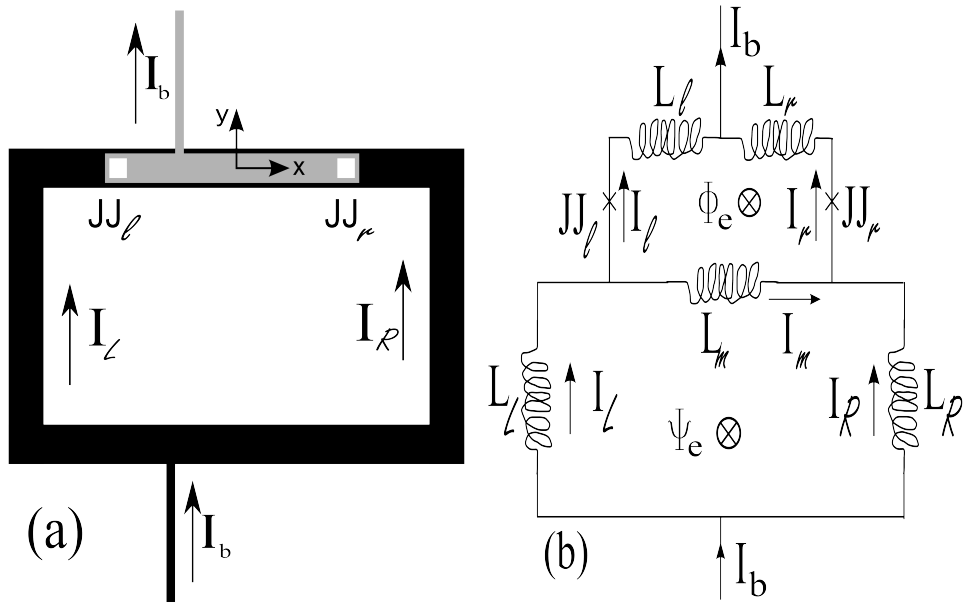}
\caption{(a) A vertical double-junction interferometer (in gray) can counts the number of flux quanta trapped in a superconducting loop (in black); (b) its equivalent circuit. $\Phi_e$ and $\Psi_e$ are two independent control variables.}
\label{fluxmeter}
\end{figure}

\begin{equation}
 n \Phi_0=\frac{\phi_\ell- \phi_r}{2\pi}\Phi_0 + \Phi_e  + L_{\ell}I_\ell- L_{m}I_m- L_{r}I_r;
\label{qua1}
\end{equation}

\noindent while, for the lower loop threated by a flux $\Psi_e$:

\begin{eqnarray}
m \Phi_0 = \Psi_e + L_{L}I_{L} + L_{m}I_{m} - L_{R}I_{R}= \nonumber \\
=\Psi_e + L_{L}(I_\ell+I_m) + L_{m}I_{m}- L_{R}(I_r-I_m)=  \nonumber \\
= \Psi_e + L_{L}I_\ell + (L_{L}+L_{m}+L_{R})I_{m}- L_{R}I_r,
\label{qua2}
\end{eqnarray}

\noindent being $I_b=I_{L}+I_{R}=I_\ell+I_r$, $I_{L}=I_\ell+I_m$ and $I_{R}=I_r-I_m$. Indeed, the lower ring might be interrupted by one or more JTSs to mimic flux qubit devices\cite{chiorescu}, however, for simplicity, we will only consider the case of a continuous loop.  Note that the magnetic fluxes $\Phi_e$ and $\Psi_e$ can be controlled separately by two independent magnetic fields applied, respectively, in the $y$ and $z$-directions. The common current $I_m$ can be eliminated form Eqs.(\ref{qua1}) and (\ref{qua2}) to yield:

\begin{eqnarray}
2\pi n=\phi_\ell- \phi_r +2\pi \phi_e + 2\pi {\sigma}(\psi_e-m) \nonumber \\
+\frac{2 \pi}{\Phi_0} [(L_{\ell}+{\sigma}L_{L})I_\ell- (L_{r}+ {\sigma}L_{R})I_r],
\label{qua3}
\end{eqnarray}

\noindent where we have introduced the dimensionless parameter ${\sigma}=L_m/L_{loop}$ with $L_{loop}=L_{L}+L_m+L_{R}$. Let's observe that, in the limits $L_m \to 0$ (${\sigma} \to 0$ - lower loop shorted) and $L_m \to \infty $ (${\sigma} \to 1$ - lower loop open), as expected, we recover Eqs.(\ref{quantize1}) and (\ref{quantize2}), respectively. Introducing the effective left and right inductances:

$$ L_{\ell}^{e}=L_{\ell}+{\sigma}L_{L} \quad\quad L_{r}^{e}=L_{r}+{\sigma}L_{R}, $$

\noindent and the corresponding effective $\beta^{}$ coefficients, $\beta_{l,r}^{e}=2 L_{\ell,r}^{e} I_{c\ell,r} /\Phi_0$, we end up with:

\begin{equation}
2\pi(n+\sigma m) =\phi_\ell- \phi_r +2\pi (\phi_e + \sigma \psi_e)
+\pi \beta_\ell^{e} \sin \phi_\ell - \pi \beta_r^{e} \sin \phi_r,
\label{qua4}
\end{equation}

\noindent i.e, with a new problem in complete correspondence to the original one stated by Eqs.(\ref{Ib}) and (\ref{quantize1}) for a simple two-junction interferometer though  with different inductances  and external flux $\phi_e + \sigma \psi_e$. From Eq.(\ref{qua4}) we see that $\sigma^{-1}$ sets the $I_c$ periodicity versus $\psi_e$ or $m$. The underlying idea is to count the trapped flux quanta $m$ by detecting the corresponding discrete changes $\Delta I_c$ in the VJI supercurrent.

\begin{figure}[htb]
\includegraphics[width=14cm]{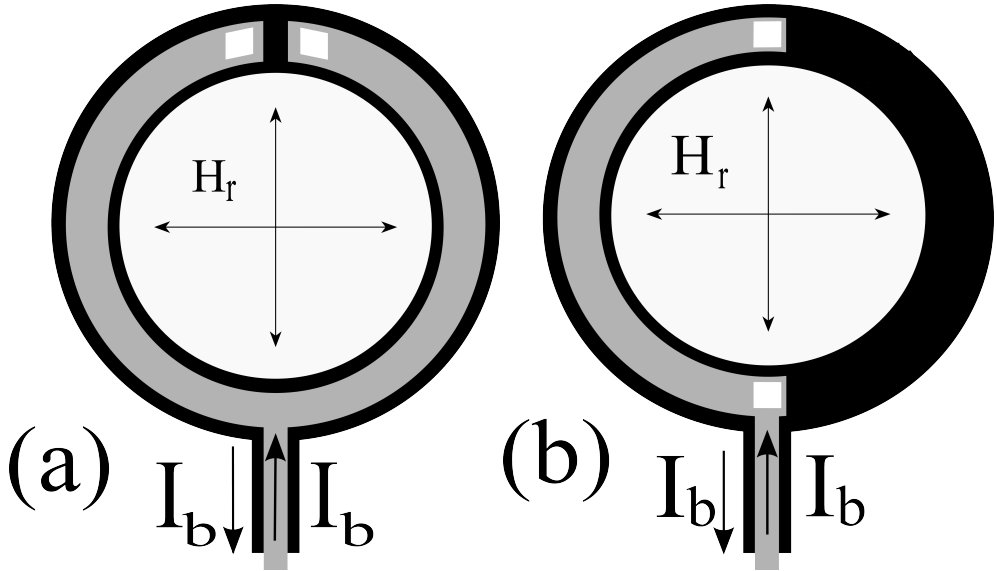}
\caption{Sketch (not in scale) of two circular vertical double junction interferometers. The base electrode is in black, the top electrode in gray and the junction area is white: (a) symmetric bias and (b) asymmetric bias.}
\label{rings}
\end{figure}

\noindent Figs.\ref{rings} show the layouts of two practical devices capable to readout the ring winding number (symmetric and asymmetric configuration). In the former case, $L_{\ell} = L_{r} \simeq L_{loop}/2$ and, since the circular VJI spans over almost 360 degree, $L_{L}=L_{R}<<L_m \simeq L_{loop}$, so that $\sigma \simeq 1$. It is worth pointing out that\cite{vanDuzer}, due to the presence of a counter electrode acting as a superconducting ground plane, the ring inductance is given by the product of its circumference times the inductance per unit length in Eq.(\ref{L}), $L_{loop}=2\pi R {\mathcal L}$, which is considerably smaller than the inductance for an isolated ring $L_{ring}=\mu_0 R (\ln 16R/w - 2)$. For the asymmetric configuration in Fig.\ref{rings}(b), $L_{\ell}<< L_{r}$ and $L_L<<L_R \simeq L_{ring}/2<L_m \simeq \pi R {\mathcal L}$ (the last inequalities guarantees the feed asymmetry). The asymmetric design provides a better sensitivity and, at the same time, allows to discriminate between flux and antiflux quanta. Of course, both $\phi_e$ and $\psi_e$ need to accurately known, otherwise one can only measure the relative changes. Let us note that the normalized flux $\phi_e$ only depends on the radial component $H_r$ of an eventual externally applied magnetic field, while $\psi_e$ is related to its transverse component $H_z$. For calibration purposes it is possible to use the $I_c(\psi_e)$ dependence when $n$, $m$ and $\phi_e$ are null (or known).

\noindent Indeed, this method is strongly inspired by the results found in investigating the spontaneous fluxoid formation in superconducting loops based on the detection of the persistent currents circulating around a hole in a superconducting film, when one or more fluxoids are trapped inside the hole\cite{PRL09}. In the most recent version of this experiments long asymmetric inline JTJs were used to sense the circulating currents and a current gain  $d I_c/ dI_b$ as large as $0.8$ has been achieved\cite{PRB11}, not far from its theoretical value\cite{note}.



\section{CONCLUDING REMARKS}

A large variety of Josephson devices capable of current amplification were the object of many research activity even before the discovery of the high-T$_c$ superconductors\cite{zegh,yoshida,pepe}. The two-junction interferometer is an excellent example of the Josephson circuitry which combines the device simplicity with great dynamic behavior. Notably, the asymmetric quantum interferometer can be viewed as a three-terminal device with current amplification. In this paper, a novel transistor-like device capable of current amplification is analyzed and a theory for its behavior is presented. In the proposed device, the interferometer supercurrent is modulated by a dissipationless current flowing in the base electrode of a superconducting strip line. Simple estimates indicate that this device can indeed be used as an amplifier, exhibiting sufficient gain to be interesting for practical applications. A potential drawback of such amplifiers is their limited dynamic range: as the flux variation in the SQUID has to be smaller than a flux quantum, the input signal must not exceed $\Phi_0/L$. Nevertheless the flexibility in design can compensate for such disadvantage. Two niche applications have been presented for which the experimental verification has been planned.



\section*{Acknowledgements}
\noindent The author thanks C. Granata, V.P. Koshelets and J. Mygind for useful suggestions and stimulating discussions.



%


\begin{thebibliography}
\bibitem{note} For an asymmetric in-line long Josephson tunnel junction, it is $I_c(H) =I_c(0)(1-H/H_c)$, with the critical current $I_c(0) = 2 J_c w \lambda_j$ and the critical field $H_c= 2 J_c \lambda_j$; assuming $H= I_a/w$, then the resulting theoretical gain is $dI_c/dI_a=1$.
\end{thebibliography}
\end{document}